\documentclass[prc,aps,onecolumn,showpacs,nofootinbib,superscriptaddress,preprint]{revtex4-1}
\usepackage{graphicx}
\usepackage{amssymb}
\usepackage{amsmath}
\usepackage[usenames]{color}

\allowdisplaybreaks

\begin{document}

\title{Amplitude reconstruction from complete electroproduction experiments and
truncated partial-wave expansions}

\author{L. Tiator}
\affiliation{Institut f\"ur Kernphysik der Universit\"at Mainz,
Johann-Joachim-Becher-Weg 45, 55099 Mainz, Germany}

\author{R. L. Workman}
\affiliation{Institute for Nuclear Studies and Department of Physics, The
George Washington University, Washington, DC 20052, USA}

\author{Y. Wunderlich}
\affiliation{Helmholtz-Institut f\"ur Strahlen- und Kernphysik der
Universit\"at Bonn, Nu{\ss}allee 14-16, 53115 Bonn, Germany}

\author{H. Haberzettl}
\affiliation{Institute for Nuclear Studies and Department of Physics, The
George Washington University, Washington, DC 20052, USA}

\date{\today}

\begin{abstract}

We compare the methods of amplitude reconstruction, for a complete
experiment and a truncated partial-wave analysis, applied to the
electroproduction of pseudoscalar mesons. We give examples which
show, in detail, how the amplitude reconstruction (observables
measured at a single energy and angle) is related to a truncated
partial-wave analysis (observables measured at a single energy and a
number of angles). A connection is made to existing data.
\end{abstract}

\pacs{25.20.Lj, 25.30.Rw, 11.80.Et, 11.55.Bq \hfill
\fbox{\textbf{File: \jobname.tex}} }

\maketitle

\section{Introduction and Motivation}

There have been numerous recent efforts to extract maximal
information, unbiased by any particular model, from experimental
pseudoscalar photoproduction data. These have included the study of
complete experiment analyses~\cite{CEA} (CEA) and truncated
partial-wave analyses~\cite{TPWA} (TPWA). Legendre analyses directly
applied to data~\cite{Leg} have the same motivation. The CEA
determines helicity or transversity amplitudes at a single energy
and angle, up to an overall (energy and angle dependent) phase.  The
TPWA introduces a cutoff to the partial-wave series, obtaining
multipoles for a fixed energy, with an overall unknown phase
dependent only on energy.

The methods used to study the photoproduction of pseudoscalar mesons
can be extended to the case of electroproduction, with the
introduction of longitudinal amplitudes associated with the incoming
virtual photon. An examination of the CEA was performed by
Dmitrasinovic, Donnelly and Gross~\cite{DDG} who considered the
required polarization measurements. They concluded that a CEA,
determining the electroproduction transversity amplitudes up to an
overall phase, was not possible with either recoil or target
polarization measurements alone, but required at least one
measurement from the other polarization set. They further concluded
that a CEA could be constructed without the need for more
complicated measurements involving both a polarized target and
recoil polarization detection. These conclusions assumed that all
structure functions could be separated in a set of measurements. As
in all such studies, it was also implicitly assumed that
measurements could be made arbitrarily precise.

Here we generalize our recent study~\cite{TPWA}
of the CEA and TPWA in photoproduction to
electroproduction. While the study in Ref.~\cite{DDG} focused on the CEA,
in practice, one desires multipole amplitudes that can be associated with
resonance contributions. These cannot be directly obtained from a complete
set of transversity amplitudes and the methods used in solving the CEA and
TPWA problems are quite different, as was discussed in detail in Ref.~\cite{TPWA}.

\begin{figure}
\begin{center}
\includegraphics[width=8.0cm]{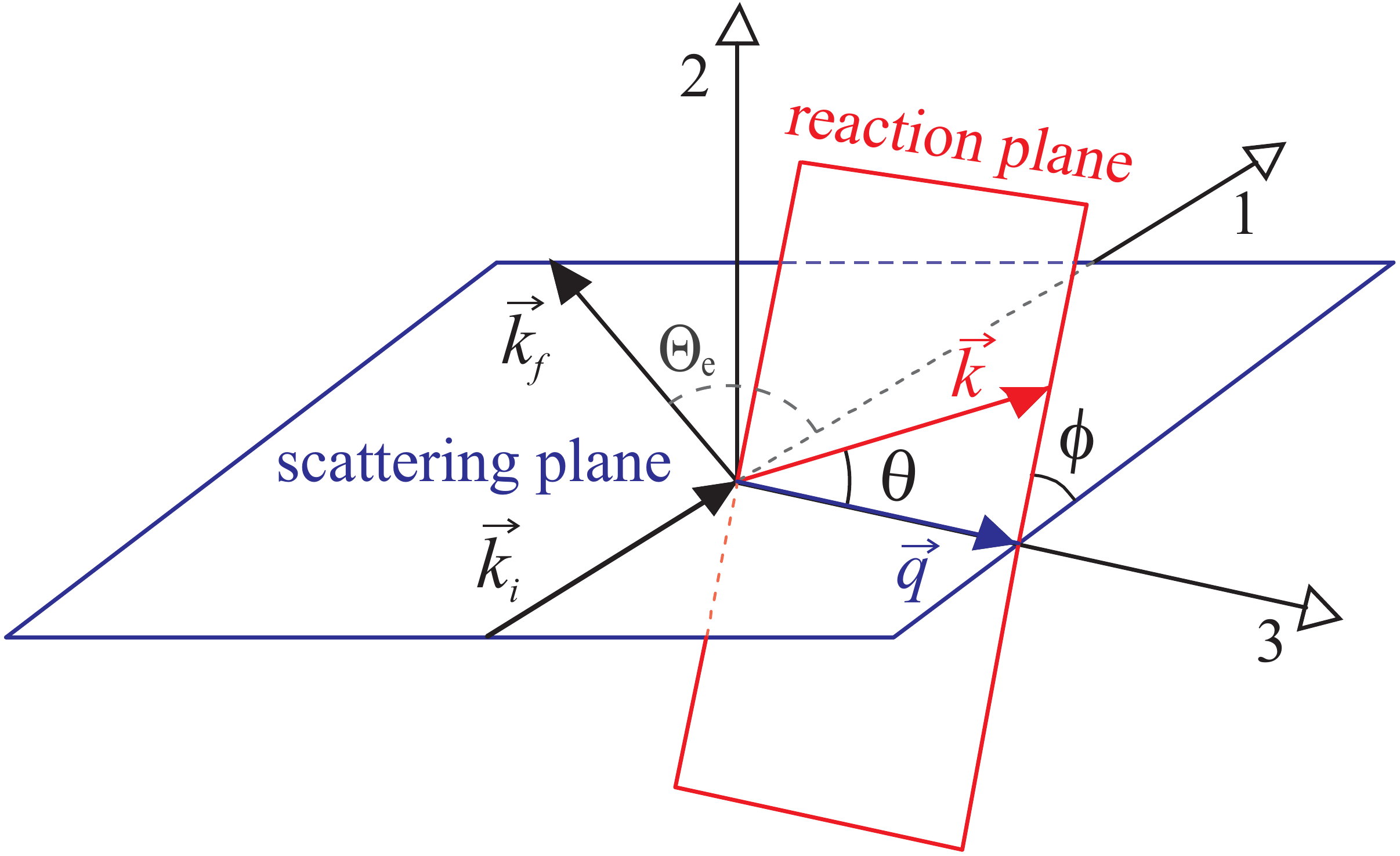}
\vspace{3mm} \caption{Kinematics of an electroproduction experiment.
The scattering plane \{1,3\} is defined by the respective incoming
and outgoing electron momenta $\vec{k}_i$, $\vec{k}_f$ with the
electron scattering angle $\Theta_e$. The reaction plane is spanned
by the virtual photon $\vec{q}$ and the outgoing meson $\vec{k}$,
scattered by the angle $\theta$. The reaction plane is tilted versus
the scattering plane by the azimuthal angle $\phi$.} \label{fig:kin}
\end{center}
\end{figure}

The electroproduction reaction, unlike photoproduction, requires detailed
knowledge of the electron scattering process producing the
interacting virtual photon. As the electron scattering and outgoing hadronic
particles define two different planes, a second angle defining their
relative orientation is required, as shown in Fig.~\ref{fig:kin}.
The virtual photon can have a non-zero value for its 4-momentum
squared, which allows for the independent variation of photon energy
and momentum.
This non-zero value also complicates the spin
structure, requiring the introduction of both longitudinal and
transverse components, as described in
Refs.~\cite{Drechsel:1992pn,Knochlein:1995qz}. Below, we first
review the electroproduction formalism. We then consider both simple
and more realistic examples of the CEA and TPWA process, showing how
the experimental requirements change.

\section{Cross Section and Polarization Degrees of Freedom}

Here we follow the notation of Ref.~\cite{Knochlein:1995qz} to
describe the pseudoscalar meson electroproduction process. As
denoted in Fig.~\ref{fig:kin}, $\Theta_e$ is the electron scattering
angle while $q$ and $k$ are the respective 4-vectors for the virtual
photon and outgoing meson, with $q^2 = \omega^2 - \bold{q}^2$,
$\omega$ and $\bold{q}$ being the photon energy and 3-momentum. The
momentum transfer is denoted by $Q^2 = -q^2$ and the ``photon
equivalent energy'' is given by $k_{\gamma}^{lab} = (W^2 -
m_i^2)/2m_i$, where $W$ is the center-of-mass energy of the hadronic
system and $m_i$ is the mass of the initial nucleon. The degree of
transverse polarization of the virtual photon is
\begin{align}
\varepsilon = \left( 1 + {2\bold{q}^2 \over {Q^2}} \tan^2 {\Theta_e
\over 2} \right)^{-1}\,,
\end{align}
with $\bold{q}$ and $\Theta_e$ expressible in either the lab or c.m.
frame. The longitudinal polarization,
\begin{align}
\varepsilon_L = {Q^2 \over {\omega^2}}\, \varepsilon\,,
\end{align}
is frame dependent.

\begin{figure}
\begin{center}
\includegraphics[width=8.0cm]{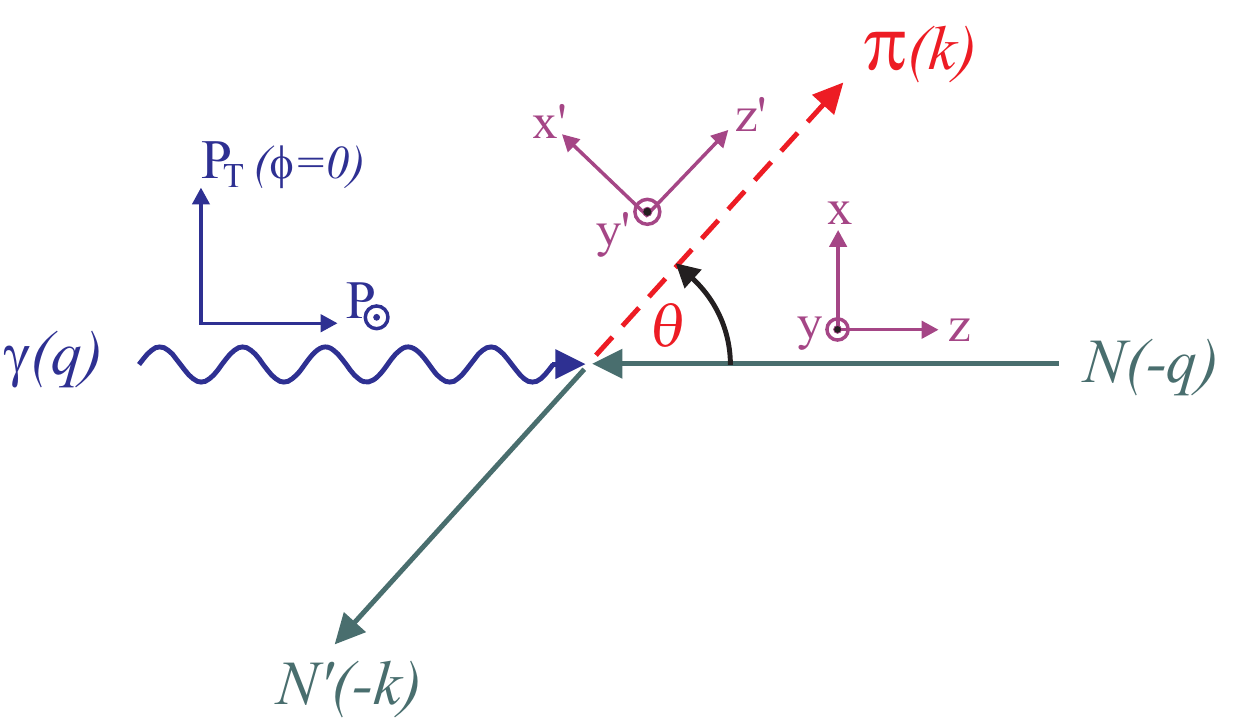} \vspace{3mm}
\caption{Frames for polarization vectors. Target and recoil
polarization are commonly defined as $\{x,y,z\}$ and $\{x',y',z'\}$
in the c.m. frame, with the $z'$ direction along the outgoing meson
$\pi(k)$. The virtual photon $\gamma(q)$ can carry different types
of polarization, including the linear and circular
polarizations, $P_T$ in the $\{x,y\}$ plane and $P_\odot$ along the
$z$ axis, as in photoproduction. In addition, the longitudinal
photon carries a polarization, $\varepsilon_L$, with further 
polarization types appearing in the $LT$ interferences of
Eq.~(\ref{eq:dsigmafull}). }\label{fig:polariz}
\end{center}
\end{figure}

Experiments with three types of polarization can be performed in
meson electroproduction: electron beam polarization, polarization of
the target nucleon and polarization of the recoil nucleon. Target
polarization will be described in the frame $\{ x, y, z \}$, with
the $z$-axis pointing in the direction of the photon momentum $\hat{
\bold{q}}$, the $y$-axis perpendicular to the reaction plane, $\hat{
\bold{y}} = \hat{ \bold{q}} \times \hat{ \bold{k}} / \sin \theta$,
where $\hat{\bold{k}}$ is the direction of the outgoing meson, and
the $x$-axis given by $\hat{\bold{x}} = \hat{ \bold{y}} \times
\hat{\bold{z}}$. For recoil polarization we will use the frame $\{
x', y', z' \}$, with the $z'$-axis defined by the momentum vector of
the outgoing meson, the $y'$-axis parallel to $\hat{\bold{y}}$, and
the $x'$-axis given by $\hat{\bold{x}}' = \hat{\bold{y}}' \times
\hat{\bold{z}}'$. These frames are displayed in
Fig.~\ref{fig:polariz}.

The most general expression for a coincidence experiment considering all
three types of polarization is
%\begin{widetext}
\begin{align}
\label{eq:dsigmafull}
\frac{d\sigma_v}{d\Omega} & = \frac{|
\vec k |}{k_{\gamma}^{cm}} P_{\alpha} P_{\beta} \Big\{ R_T^{\beta
\alpha}
+ \varepsilon_L R_L^{\beta \alpha}
\nonumber \\
&\quad\mbox{}
 + \left[ 2 \varepsilon_L \left( 1 + \varepsilon \right)
\right]^{1/2} ( ^c \! R_{LT}^{\beta \alpha} \cos \phi + ^s \!
\! R_{LT}^{\beta \alpha} \sin \phi)
\nonumber \\
&\quad\mbox{}
      + \varepsilon ( ^c \! R_{TT}^{\beta \alpha} \cos 2 \phi +
^s \! R_{TT}^{\beta \alpha} \sin 2 \phi)
\nonumber \\
&\quad\mbox{}
 + h \left[ 2 \varepsilon_L ( 1 - \varepsilon ) \right]^{1/2} (
^c \! R_{LT'}^{\beta \alpha} \cos \phi + ^s \! R_{LT'}^{\beta
\alpha} \sin \phi)
\nonumber \\
&\quad\mbox{}
 + h ( 1 - \varepsilon^2 )^{1/2} R_{TT'}^{\beta \alpha} \Big\}\,,
\end{align}
%\end{widetext}
where $h$ is the helicity of the incoming electron, $P_{\alpha} =
(1, \vec P)_{\alpha}$ and $P_{\beta} = (1, \vec P')_{\beta}$. Here
$\vec P = (P_x, P_y, P_z)$ denotes the target and $\vec P' =
(P_{x'}, P_{y'}, P_{z'})$ the recoil polarization vector. The zero
components, $P_0 = 1$, lead to contributions in the cross section
which are present in the polarized as well as the unpolarized case.
In an experiment without target and recoil polarization, $\alpha =
\beta = 0$ and the only remaining contributions are $R_i^{00}$. The
functions $R_i^{\beta \alpha}$ describe the response of the hadronic
system in the process. Summation over Greek indices (0,1,2,3) is
implied. An additional superscript $s$ or $c$ on the left indicates
a sine or cosine dependence of the respective contribution on the
azimuthal angle. Some response functions vanish identically (see
Table~\ref{tab:obs1} of Ref.~\cite{Knochlein:1995qz} for a
systematic overview). The number of different response functions is
further reduced by equalities, as shown in Table~\ref{tab:obs1}, and
in the most general electroproduction experiment, 36 polarization
observables can be determined. The response functions $R^{\beta
\alpha}_i$ are real or imaginary parts of bilinear forms of the
CGLN~\cite{CGLN} amplitudes depending on the scattering angle
$\theta$.

\section{Amplitudes used in pseudoscalar meson electroproduction}

Before comparing the CEA and TPWA approaches, we continue with a
review of notation used for the underlying amplitudes.
The multipoles and CGLN~\cite{CGLN} $F$-amplitudes are related by
\begin{subequations}
\begin{align}
F_1 &= \sum_{\ell \ge 0}\left[ \left( \ell M_{\ell +} + E_{\ell +}
\right) P'_{\ell + 1} +\left( (\ell + 1) M_{\ell -} + E_{\ell -}
\right) P'_{\ell - 1} \right]\,,
\\
F_2 &= \sum_{\ell \ge 1}\left[ (\ell + 1) M_{\ell +} + \ell M_{\ell
-} \right]  P'_{\ell }\,,
\\
F_3 &= \sum_{\ell \ge 1}\left[ \left( E_{\ell +} - M_{\ell +}
\right) P''_{\ell + 1} +\left( E_{\ell -} + M_{\ell -} \right)
P''_{\ell - 1} \right]\,,
\\
F_4 &= \sum_{\ell \ge 2}\left[ M_{\ell +} - E_{\ell +} -  M_{\ell -}
- E_{\ell -} \right]  P''_{\ell }\,,
\\
F_5 &= \sum_{\ell \ge 0} \left[ (\ell + 1) L_{\ell +} P'_{\ell +1}
-\ell \; L_{\ell -} P'_{\ell - 1} \right]\,,
\\
F_6 &= \sum_{\ell \ge 1} \left[ \ell \; L_{\ell - } - (\ell + 1)
L_{\ell +} \right] P'_{\ell}\,.
\end{align}
\end{subequations}
The definition of helicity amplitudes is subject to phase
conventions. Here, we choose the conventions of~\cite{Jacob}, which
were also used by Walker in~\cite{Walker} for photoproduction.
Without loss of generality, we set $\phi=0$,
\begin{subequations}
\begin{align}
H_1 &= -\frac{1}{\sqrt{2}} \sin \theta \cos \frac{\theta}{2} ( F_3 +
F_4 )\,,
\\
H_2 &= \sqrt{2} \cos \frac{\theta}{2} \left( F_2 - F_1 + ( F_3 - F_4
) \sin^2 \frac{\theta}{2} \right)\,,
\\
H_3 &= \frac{1}{\sqrt{2}} \sin \theta \sin \frac{\theta}{2} (F_3 -
F_4)\,,
\\
H_4 &= \sqrt{2} \sin \frac{\theta}{2} \left(F_1 + F_2 + (F_3 + F_4)
\cos^2 \frac{\theta}{2} \right)\,,
\\
H_5 &= \cos \frac{\theta}{2} (F_5 + F_6 ) \,,
\\
H_6 &= -\sin \frac{\theta}{2} (F_5 - F_6) \,.
\end{align}
\end{subequations}
Finally, transversity amplitudes can be constructed~\cite{bds,DDG}
from these helicity amplitudes,
\begin{subequations}
\begin{align}
b_1 &= \frac{1}{2} \left[ \left( H_1 + H_4 \right) \; + \; i \;
\left( H_2 - H_3 \right) \right]\,,
 \\
b_2 &= \frac{1}{2} \left[ \left( H_1 + H_4 \right) \; - \; i \;
\left( H_2 - H_3 \right) \right]\,,
 \\
b_3 &= \frac{1}{2} \left[ \left( H_1 - H_4 \right) \; - \; i \;
\left( H_2 + H_3 \right) \right]\,,
 \\
b_4 &= \frac{1}{2} \left[ \left( H_1 - H_4 \right) \; + \; i \;
\left( H_2 + H_3 \right) \right]\,,
\\
b_5 &= \frac{1}{\sqrt{2}} \left[ H_5 \; + \; i \; H_6 \right]\,, \\
b_6 &= \frac{1}{\sqrt{2}} \left[ H_5 \; - \; i \; H_6 \right]\,.
\end{align}
\end{subequations}
Here we note that the definitions of both helicity and transversity
amplitudes are not unique. Apart from phase conventions, 
different numbering choices can also be found in the literature. Here we
follow the definitions of Barker et al.~\cite{bds}. In
Table~\ref{tab:obs1}, expressions for the response functions,
appearing in Eq.~(\ref{eq:dsigmafull}), are given in terms of both
the helicity and transversity amplitudes.
In the following, we will suppress the superscripts $c$ and $s$ for
interference terms. As can be seen in Table~\ref{tab:obs1}, for a
specific polarization, the assignment of this superscript 
is always unique.

%  Table I
\begin{table*}
\begin{center}
\caption{\label{tab:obs1}Spin observables expressed in terms of
helicity and transversity amplitudes. Also listed are alternate
(ALT) observables, differing by at most a sign in their definition,
and associated photoproduction observables ($\gamma$). Note, that
these expressions are not uniquely defined. We follow the
conventions of Refs.~\cite{bds,Walker}. }
\bigskip
\begin{tabular}{|c|c|c|c|c|}
\hline
 Obs & ALT & $\gamma$  &  Helicity                                             & Transversity    \\[-1.7ex]
     &  &       &  representation                                       & representation  \\
\hline
$R_T^{00}$ & $-^c \! R_{TT}^{y'y}$ & $I$ & $\frac{1}{2}$ ($|H_1|^2 + |H_2|^2 + |H_3|^2 + |H_4|^2$)  & $\frac{1}{2} (|b_1|^2+|b_2|^2+|b_3|^2+|b_4|^2$)   \\
$R_T^{0y}$ & $-^c \! R_{TT}^{y'0}$ & $\check{T}$ &$-$Im$(H_2 H_1^* + H_4 H_3^*)$                           & $\frac{1}{2}(|b_1|^2-|b_2|^2-|b_3|^2+|b_4|^2)$      \\
$R_T^{y'0}$ & $-^c \! R_{TT}^{0y}$ & $\check{P}$& Im$(H_3 H_1^* + H_4 H_2^*)$                           & $\frac{1}{2}(|b_1|^2-|b_2|^2+|b_3|^2-|b_4|^2$)   \\
$R_T^{x'x}$ & $-^c \! R_{TT}^{z'z}$& $\check{T}_{x'}$ & Re$(H_4 H_1^* + H_3 H_2^*)$                        & Re($b_1 b_2^* - b_4 b_3^*$)       \\
           $R_T^{x'z}$&$ ^c \! R_{TT}^{z'x}$&$-\check{L}_{x'}$      & Re$(H_3 H_1^* - H_4 H_2^*)$                        & ${\rm Im}(b_4 b_3^* - b_1 b_2^*)$       \\
           $R_T^{z'x}$&$ ^c \! R_{TT}^{x'z}$&$\check{T}_{z'}$      & Re$(H_2 H_1^* - H_4 H_3^*)$                        & ${\rm Im}(b_1 b_2^* + b_4 b_3^*)$     \\
           $R_T^{z'z}$&$-^c \! R_{TT}^{x'x}$&$\check{L}_{z'}$      & $\frac{1}{2}(|H_1|^2 - |H_2|^2 - |H_3|^2 + |H_4|^2)$ & ${\rm Re}(b_1 b_2^* + b_4 b_3^*)$     \\
\hline
$R_L^{00}$ & $-R_L^{y'y}$ && $|H_5|^2 + |H_6|^2$                           & $|b_5|^2 + |b_6|^2$     \\
$R_L^{0y}$ & $-R_L^{y'0}$ && $-2{\rm Im}(H_6 H_5^*)$                        & $|b_5|^2 - |b_6|^2$     \\
$R_L^{x'x}$ & $-R_L^{z'z}$ && $-|H_5|^2 + |H_6|^2$                           & $-2{\rm Re}(b_6 b_5^*)$    \\
      $R_L^{z'x} $ & $R_L^{x'z} $ && $2{\rm Re}(H_6 H_5^*)$                        & $-2{\rm Im}(b_6 b_5^*)$      \\
\hline
$^c \! R_{LT}^{00}$ & $-^c \! R_{LT}^{y'y}$ && $\frac{1}{\sqrt{2}}{\rm Re}\left((H_1 - H_4) H_5^* + (H_2 + H_3) H_6^*\right)$  & ${\rm Re}(b_6 b_3^* + b_5 b_4^*)$   \\
$^s \! R_{LT}^{0x}$ & $ ^c \! R_{LT'}^{y'z}$ && $\frac{1}{\sqrt{2}}{\rm Im}((H_3 - H_2) H_5^* - (H_1 + H_4) H_6^*)$ & ${\rm Re}(b_1 b_6^* - b_5 b_2^*)$  \\
      $^c \! R_{LT}^{0y} $ & $-^c \! R_{LT}^{y'0}$ && $-\frac{1}{\sqrt{2}}{\rm Im}((H_2 + H_3) H_5^* - (H_1 - H_4)H_6^*)$   & ${\rm Re}(b_5 b_4^* - b_6 b_3^*)$    \\
      $^s \! R_{LT}^{0z} $ & $-^c \! R_{LT'}^{y'x}$ && $-\frac{1}{\sqrt{2}}{\rm Im}((H_1 + H_4) H_5^* - (H_2 - H_3) H_6^*)$   & ${\rm Im}(b_5 b_2^* - b_1 b_6^*)$    \\
      $^s \! R_{LT}^{x'0}$ & $-^c \! R_{LT'}^{z'y}$ && $\frac{1}{\sqrt{2}}{\rm Im}((H_2 - H_3) H_5^* - ( H_1 + H_4) H_6^*)$  & ${\rm Re}(b_6 b_2^* - b_1 b_5^*)$         \\
      $^s \! R_{LT}^{z'0}$ & $ ^c \! R_{LT'}^{x'y}$ && $-\frac{1}{\sqrt{2}}{\rm Im}((H_1 + H_4) H_5^* + (H_2 - H_3) H_6^*)$   & ${\rm Im}(b_6 b_2^* - b_1 b_5^*)$ \\
      $^c \! R_{LT}^{x'x}$ & $-^c \! R_{LT}^{z'z}$  && $-\frac{1}{\sqrt{2}}{\rm Re}((H_1 - H_4) H_5^* - (H_2 + H_3) H_6^*)$     &  $-{\rm Re}(b_5 b_3^* + b_6 b_4^*)$ \\
      $^c \! R_{LT}^{z'x}$ & $ ^c \! R_{LT}^{x'z}$  && $\frac{1}{\sqrt{2}}{\rm Re}((H_2 + H_3 )H_5^*  + (H_1 - H_4) H_6^*)$     & ${\rm Im}(b_5 b_3^* - b_6 b_4^*)$   \\
\hline
\end{tabular}
\end{center}
\end{table*}
%  Table I continued
\setcounter{table}{0}
\begin{table*}
\begin{center}
\caption{\label{tab:obs2} (continued)
}
\bigskip
\begin{tabular}{|c|c|c|c|c|}
\hline
 Obs & ALT & $\gamma$  &  Helicity                                             & Transversity    \\[-1.7ex]
     &  &       &  representation                                       & representation  \\
\hline
           $^c \! R_{TT}^{00}$&$-R_{T}^{y'y}$&$-\check{\Sigma}$  & ${\rm Re}(H_3 H_2^* - H_4 H_1^*)$    &  $\frac{1}{2}(-|b_1|^2 - |b_2|^2 + |b_3|^2 + |b_4|^2)$ \\
           $^s \! R_{TT}^{0x}$&$ R_{TT'}^{y'z}$&$\check{H}$  & ${\rm Im}(H_3 H_1^* - H_4 H_2^*)$    & ${\rm Re}(b_1 b_3^* - b_4 b_2^*)$      \\
           $^s \! R_{TT}^{0z}$&$-R_{TT'}^{y'x}$&$-\check{G}$   & $-{\rm Im}(H_4 H_1^* + H_3 H_2^*)$  & ${\rm Im}(b_4 b_2^* - b_1 b_3^*)$   \\
           $^s \! R_{TT}^{x'0}$&$-R_{TT'}^{z'y}$&$\check{O}_x$  & ${\rm Im}(H_2 H_1^* - H_4 H_3^*)$  &  ${\rm Re}(b_3 b_2^* - b_1 b_4^*)$     \\
           $^s \! R_{TT}^{z'0}$&$ R_{TT'}^{x'y}$&$\check{O}_z$  & ${\rm Im}(H_3 H_2^* - H_4 H_1^*)$  & ${\rm Im}(b_3 b_2^* - b_1 b_4^*)$       \\
\hline
           $^s \! R_{LT'}^{00}$&$-^s \! R_{LT'}^{y'y}$& & $- \frac{1}{\sqrt{2}}{\rm Im}((H_1 - H_4) H_5^* + (H_2 + H_3)H_6^*)$     & ${\rm Im}(b_6 b_3^* + b_5 b_4^*)$     \\
           $^c \! R_{LT'}^{0x}$&$-^s \! R_{LT}^{y'z}$& & $\frac{1}{\sqrt{2}}{\rm Re}((H_2 - H_3) H_5^* + (H_1 + H_4) H_6^*)$ & ${\rm Im}(b_1 b_6^* + b_5 b_2^*)$     \\
           $^s \! R_{LT'}^{0y}$&$-^s \! R_{LT'}^{y'0}$& & $- \frac{1}{\sqrt{2}}{\rm Re}((H_2 + H_3) H_5^* + (H_4 - H_1) H_6^*)$  & ${\rm Im}(b_5 b_4^* - b_6 b_3^*)$     \\
      $^c \! R_{LT'}^{0z}$ &$ ^s \! R_{LT}^{y'x}$&  & $\frac{1}{\sqrt{2}}{\rm Re}((H_1 + H_4)H_5^* + (H_3 - H_2) H_6^*)$  & ${\rm Re}(b_1 b_6^* + b_5 b_2^*)$     \\
      $^c \! R_{LT'}^{x'0}$&$ ^s \! R_{LT}^{z'y}$& & $\frac{1}{\sqrt{2}}{\rm Re}((H_3 - H_2) H_5^* + (H_1 +H_4) H_6^*)$   & $-{\rm Im}(b_1 b_5^* + b_6 b_2^*)$    \\
      $^c \! R_{LT'}^{z'0}$&$-^s \! R_{LT}^{x'y}$& & $\frac{1}{\sqrt{2}}{\rm Re}((H_1 + H_4) H_5^* + (H_2 - H_3) H_6^*)$ & ${\rm Re}(b_1 b_5^* + b_6 b_2^*)$      \\
      $^s \! R_{LT'}^{x'x}$&$-^s \! R_{LT'}^{z'z}$& & $\frac{1}{\sqrt{2}}{\rm Im}((H_1 - H_4) H_5^* - (H_2 + H_3) H_6^*)$  & $-{\rm Im}(b_5 b_3^* + b_6 b_4^*)$   \\
      $^s \! R_{LT'}^{z'x}$&$ ^s \! R_{LT'}^{x'z}$& & $- \frac{1}{\sqrt{2}}{\rm Im}((H_2 + H_3) H_5^* + (H_1 - H_4) H_6^*)$ & ${\rm Re}(b_6 b_4^* - b_5 b_3^*)$  \\
\hline
      $ R_{TT'}^{0x}$ &$-^s \! R_{TT}^{y'z}$&$\check{F} $  &  ${\rm Re}(H_2 H_1^* + H_4 H_3^*)$   & ${\rm Im}(b_1 b_3^* + b_4 b_2^*)$    \\
      $ R_{TT'}^{0z}$ &$ ^s \! R_{TT}^{y'x}$&$-\check{E} $  &  $\frac{1}{2}(|H_1|^2 - |H_2|^2 + |H_3|^2 - |H_4|^2) $   & ${\rm Re}(b_1 b_3^* + b_4 b_2^*)$    \\
      $ R_{TT'}^{x'0}$&$ ^s \! R_{TT}^{z'y}$&$-\check{C}_{x'} $ &  ${\rm Re}(H_3 H_1^* + H_4 H_2^*)$  & $- {\rm Im}(b_1 b_4^* + b_3 b_2^*)$         \\
      $ R_{TT'}^{z'0}$&$-^s \! R_{TT}^{x'y}$&$-\check{C}_{z'} $ & $\frac{1}{2}(|H_1|^2 + |H_2|^2 - |H_3|^2 - |H_4|^2)$   & ${\rm Re}(b_1 b_4^* + b_3 b_2^*)$ \\
\hline
\end{tabular}
\end{center}
\end{table*}

Transversity amplitudes often simplify the discussion of amplitude
reconstruction in photoproduction, as the unpolarized and single-polarization
observables determine their moduli. Another
simplification is the property
\begin{equation}
b_2(\theta) = -b_1(- \theta)\,,\quad  b_4(\theta) = -b_3(-
\theta)\,,
  \;{\rm and}\quad~b_6(\theta ) = b_5(- \theta)\,,
\label{eq:TraAmplsSymmetryRelations}
\end{equation}
which allows one to parameterize only three of the six transversity
amplitudes. The form introduced by Omelaenko~\cite{omel},
\begin{subequations}\label{eq:BallOriginalOmelDecomposition}
\begin{align}
b_1 &= c\,a_{2L} {{e^{i\theta /2} }\over {(1+x^2)^L}}
\prod_{i=1}^{2L} (x-\alpha_i)\,,
\label{eq:B1OriginalOmelDecomposition}
\\
b_3 &= -c\,a_{2L} {{e^{i\theta /2} }\over {(1+x^2)^L}}
\prod_{i=1}^{2L} (x-\beta_i)\,,
\label{eq:B3OriginalOmelDecomposition}
\end{align}
\end{subequations}
with $x=\tan ( \theta /2 )$ and $L$ being the upper limit for $\ell$, is
convenient for a truncated partial-wave analysis, as the ambiguities can be
linked to the conjugation of the complex roots of the above relations, with a
constraint
\begin{equation}
\prod_{i=1}^{2L} \alpha_i \; = \; \prod_{i=1}^{2L} \beta_i~.
\label{eq:OriginalOmelConstraint}
\end{equation}
The quantity $c$ is a constant and $a_{2L}$ is proportional to the backward photoproduction cross
section~\cite{TPWA,omel}.

For the amplitudes $b_{5}$ and $b_{6}$, which are present in
electroproduction in addition to the four transverse amplitudes, it
is feasible to write a linear-factor decomposition according to
Omelaenko, similar to expressions
(\ref{eq:B1OriginalOmelDecomposition}) and
(\ref{eq:B3OriginalOmelDecomposition}). As the resulting
non-redundant transversity amplitude, we pick here $b_{6}$ and the
expression is
\begin{equation}
b_6 = c\,d_{2L} {{e^{ i\theta /2} }\over {(1+x^2)^L}}
\prod_{i=1}^{2L} (x-\gamma_i)~.
\label{eq:B5ElectroproductionOmelDecomposition}
\end{equation}
The amplitude $b_{5}$ is then specified via the constraint given in
(\ref{eq:TraAmplsSymmetryRelations}). The $2L$ complex roots
$\gamma_{i}$ determine the purely longitudinal amplitudes $b_{5}$
and $b_{6}$, while the constant $c$ is the same as in
(\ref{eq:B1OriginalOmelDecomposition}) and
(\ref{eq:B3OriginalOmelDecomposition}). The quantity $d_{2L}$ is
another polynomial normalization coefficient, which may differ from
$a_{2L}$.

However, no constraint among the $\gamma$-roots has been found which
would be analogous to Omelaenko's relation
(\ref{eq:OriginalOmelConstraint}) for the $\alpha$- and
$\beta$-roots and we conjecture that no such additional constraint
for the $\gamma_{i}$ exists. This may be substantiated by the fact
that the number of real degrees of freedom for the parameterizations
of $b_{5}$ and $b_{6}$ in terms of multipoles, as well as in terms
of roots, exactly match.

For every truncation order $L$, one has $2L+1$ complex longitudinal
multipoles, i.e. the $S$-wave $L_{0+}$ and two new multipoles
$L_{\ell \pm}$ for every new order in $\ell$. This corresponds in
terms of mulipoles to $4 L +2$ real degrees of freedom. In terms of
roots, one has the $\gamma_{i}$ which comprise a set of $2L$ complex
variables or $4L$ real degrees of freedom. In addition to this, the
complex normalization coefficient $d_{2L}$ also defines $b_5$ and
$b_6$, which brings the total number of real variables to $4L+2$ in
this case as well.

The only issue not considered until now is the overall phase, either
of (for instance) $L_{0+}$, in case of the
multipole-parametrization, or $d_{2L}$ in case of roots, which
remains undetermined if only longitudinal observables are measured.
This would reduce the number of real degrees of freedom by one.
However, in electroproduction, the mixed observables of type $LT$
can very well fix this overall phase, leaving the unknown phase
information in one of the quantities specifying the purely
transverse amplitudes, e.g. $E_{0+}$. Therefore, the number $4L+2$
real variables for longitudinal multipoles remains true for the most
general case in electroproduction.

For the transverse multipoles, the situation is the same as in
photoproduction with $4L$ multipoles, i.e. the $S$-wave $E_{0+}$,
the $P$-waves $E_{1+},M_{1+},M_{1-}$ and four new multipoles
$E_{\ell\pm},M_{\ell\pm}$ for every new order in $\ell$. If we
subtract the overall free phase, which is typically assumed for the
$E_{0+}$ multipole, we have $8L-1$ real values to be determined by
the experiment.

Altogether with longitudinal and transverse multipoles, the most
general case in electroproduction is described by $6L+1$ $E,M,L$
multipoles, and $12L+1$ real values have to be determined by the
experiment. And one of those, e.g. $E_{0+}$, can be chosen to be
positive.

\section{Complete Experiment Analysis (CEA)}

In electroproduction, the CEA needs to determine six complex
amplitudes at a given energy and angle, e.g. helicity amplitudes
$H_{1,\ldots, 6}$ or transversity amplitudes $b_{1,\ldots, 6}$ up to
an overall phase, which is naturally also energy and angle
dependent. This requires the determination of 11 real numbers, where
one of them can be chosen to be positive. In principle this could
work with 11 observables, but due to quadrant ambiguities, a minimum
of 12 will be generally required.

Choosing 12 observables out of 36 will allow more than a billion
different sets. Even restricting to meaningful sets, including
transverse, longitudinal and $LT$ interference terms, still gives
millions of non-trivial sets that need to be checked for
completeness.

Two strategies seem to work straightforwardly. First, one would
select the six observables that are defined only by moduli of
transversity amplitudes,
$R_T^{00},R_T^{0y},R_T^{y'0},R_L^{00},R_L^{0y},R_{TT}^{00}$. Then
five relative angles need to be defined from six out of the
remaining 30 interference terms. Even if thousands of such sets will
lead to complete sets of 12 observables, it is not obvious how these
observables should be chosen. As can be seen in
Table~\ref{tab:obs1}, except for $b_5 b_6^*$, all interference terms
appear as linear combinations, e.g. $b_1 b_2^*\pm b_3 b_4^*$ and a
direct separation would always require a measurement of both $\pm$
combinations. Therefore, a separation of 5 angles as cosine and sine
functions would naively require 10 observables, leading altogether
to 16, and it is nontrivial to reduce this number by four
observables to find the minimum number of eight.

A second approach is to start with a complete set of 8 observables
for the transverse amplitudes $b_1,b_2,b_3,b_4$ in a CEA of
photoproduction. Such studies are also nontrivial, but have been
intensively studied in the literature, and the most comprehensive
study was done by Chiang and Tabakin~\cite{CEA}. Having chosen any
of almost 4500 possible complete sets of 8 observables leads to a
unique determination of four moduli and 3 relative angles. Then with
four additional $LT$ interference terms, such as $\mbox{Re}(b_1 b_5^*
\pm b_2 b_6^*)$  and $\mbox{Im}(b_1 b_5^* \pm b_2 b_6^*)$, the remaining
moduli $|b_5|,|b_6|$ and the relative phases of $b_5$ and $b_6$ to
the already known transverse amplitudes $b_1,b_2$ are uniquely
determined. This leads to, for example, the complete set of 12 observables
$R_{T}^{00},R_{T}^{0y},R_{T}^{y0},R_{TT}^{00},R_{TT}^{0x},R_{TT'}^{0x},
R_{TT}^{z'0},R_{TT'}^{z'0},R_{LT}^{x'0},R_{LT}^{z'0},R_{LT'}^{x'0},R_{LT'}^{z'0}$.
In this case four $LT$ interference terms with beam-recoil
polarization have been used.

Alternatively, another three combinations can be chosen with $b_2
b_5^* \pm b_1 b_6^*$, $b_3 b_5^* \pm b_4 b_6^*$ and $b_4 b_5^* \pm
b_3 b_6^*$. Looking at Table~\ref{tab:obs1}, one finds that the
first set, $b_1 b_5^* \pm b_2 b_6^*$, requires recoil polarization,
the second one, $b_2 b_5^* \pm b_1 b_6^*$, target polarization and
the third one, $b_3 b_5^* \pm b_4 b_6^*$, would even require both
target and recoil polarization. The last one, $b_4 b_5^* \pm b_3
b_6^*$, corresponds to the observables
$R_{LT}^{00},R_{LT}^{0y},R_{LT'}^{00},R_{LT'}^{0y}$ which is
identical to $R_{LT}^{00},R_{LT}^{y'0},R_{LT'}^{00},R_{LT'}^{y'0}$
and can therefore be measured with either target or recoil
polarization.

By this rather simple strategy, we have already found four times the
number of possible complete photoproduction sets, which amounts to
almost 18000 complete sets of electroproduction.

Using the Mathematica NSolve function and integer algebra for
randomly chosen real and imaginary parts of amplitudes, we can test
any given set of 12 observables for completeness. Given the enormous
number of possibilities with hundreds of millions of sets with 12
observables (where only $R_T^{00}$ is set), we have not yet
performed a systematic search for all possible complete sets as 
was done for photoproduction in our previous work~\cite{TPWA}.

\section{Amplitude Reconstruction}

\subsection{\boldmath Simplest case: $L = 0$}

In photoproduction this case is trivial, involving only a single
multipole amplitude. Here, in Set 1 of Table~\ref{tab3}, there are
two multipoles ($E_{0+}$ and $L_{0+}$), producing two independent
helicity or transversity amplitudes, requiring only 3 measurements
(e.g. $R_T^{00}$, $R_{LT}^{0y}$, $R_{LT'}^{0y}$) at a single energy
and angle, which solves both the CEA and TPWA. This is a special
case, where the absolute squares of the two multipoles are not mixed
together, but already separated in $R_T^{00}=|E_{0+}|^2$ and
$R_L^{00}=|L_{0+}|^2$. Therefore, $R_T^{00}$ gives directly the
$E_{0+}$ multipole, which can freely be taken with a positive value,
and for the absolute value $|L_{0+}|$ and the relative angle, the
two selected $LT$ interference terms are sufficient.

It should be noted, however, that in principle, through the
Rosenbluth separation of $R_T$ and $R_L$, the determination of $R_T$
gives also $R_L$, and therefore the three observable case is
essentially academic; in practice a fourth measurement
needs to be done. We will return to this Rosenbluth issue later on.

\subsection{\boldmath Case: $J= 1/2$}

Here, in Set 2 of Table~\ref{tab3}, there are four multipoles
involved ($E_{0+}$, $M_{1-}$, $L_{0+}$, $L_{1-}$) producing four
independent helicity or transversity amplitudes. The separation into
longitudinal and transverse pairs suggests two strategies for
finding a complete set of eight measurements for a CEA in this case.
Sets of four observables would determine either the transverse or
longitudinal pairs, up to an overall phase, but would leave the
relative phase between the pairs undetermined. One method: Take the
set of four measurements determining ($E_{0+}$ and $M_{1-}$) up to
an overall phase ($R_T^{00}$, $R_T^{y'0}$, $R_T^{x'z}$,
$R_T^{z'z}$). Add to this a set of four measurements defining the
relative phases of $L_{0+}$ and $L_{1-}$ to $E_{0+}$ and $M_{1-}$
respectively ($R_{LT}^{0y}$, $R_{LT}^{x'x}$, $R_{LT}^{z'0}$,
$R_{LT'}^{0y}$). Second method: Take the sets of four measurements
defining the longitudinal and transverse pairs up to an overall
phase. Remove one measurement from each set and replace with a pair
of interference terms. This leads, for example, to the set
($R_T^{00}$, $R_T^{y'0}$, $R_T^{z'z}$, $R_L^{00}$, $R_L^{0y}$,
$R_L^{z'x}$, $R_{LT}^{00}$, $R_{LT'}^{00}$).

Furthermore, longitudinal observables $R^{\beta\alpha}_L$ can be
avoided by getting the same information from $LT$ interference
terms, and a solution is found with a minimum number of five
observables, with some of these measured at two angles.

As a general rule, for $n$ complex multipoles we need $2n$
independent measurements. Due to the free overall phase (we always
assume $E_{0+}$ real and positive), there are $2n-1$ free
parameters. However, in order to solve the quadrant ambiguity, we
generally need one more measurement. In the special case of $L=0$
(Set 1) this was not needed but, as was mentioned, this case is
exceptional.

%  Table II
\begin{table*}[t!]
\caption{\label{tab3}Examples of measurements at a single energy for
CEA and TPWA. The number of different measurements (n), different
observables (m) and different angles (k) needed for a complete
analysis are given as $n(m)k$. Entries with a $\dagger$ do not allow
the comparison CEA $\leftrightarrow$ TPWA. For cases with only one
angle, the CEA and TPWA are equivalent. The number of necessary
distinct angular measurements is given in brackets.
\\}
\begin{tabular}{|c|l|c|c|l|}
\hline Set & Included Partial Waves & CEA & TPWA & Complete Sets for TPWA\\
\hline
1 &$L=0\; (E_{0+},L_{0+})$                   & 3(3) & 3(3)1 & $R_T^{00}[1],R_{LT}^{0y}[1],R_{LT'}^{0y}[1]$\\
  & 2 $S$ wave multipoles                    &      &   & \\
\hline
2 &$J=1/2\; (E_{0+},M_{1-},L_{0+},L_{1-})$   & 8(8) & 8(8)1 & $R_T^{00}[1],R_T^{y'0}[1],R_T^{z'z}[1],R_L^{00}[1],R_L^{0y}[1],R_{L}^{z'x}[1],$\\
  &  4 $S,P$ wave multipoles                 &      &       & $\; R_{LT}^{00}[1],R_{LT'}^{00}[1]$\\
  &                                          & 8(8) & 8(8)1 & $R_T^{00}[1],R_T^{y'0}[1],R_T^{x'z}[1],R_T^{z'z}[1],R_{LT}^{0y}[1],R_{LT}^{0z}[1],$\\
  &                                          &      &       & $\; R_{LT}^{x'x}[1],R_{LT'}^{0y}[1]$\\
  &                                          &      & 8(5)2 & $R_T^{00}[2],R_T^{y'0}[1],R_{LT}^{00}[1],R_{LT}^{0x}[2],R_{LT'}^{0x}[2]$\\
 \hline
3 &$L=0,1\; (L_{0+},L_{1-},L_{1+})$          & $\dagger$  &    &  TPWA at 1 angle not possible\\
  & full set of 3 longitudinal               &   & 7(4)2 & $R_L^{00}[2],R_{L}^{0y}[2],R_{L}^{x'x}[1],R_{L}^{z'x}[2]$\\
  & $S,P$ wave multipoles                    &   & 6(3)3 & $R_L^{00}[3],R_{L}^{0y}[2],R_{L}^{x'x}[1]$\\
 \hline
4 &$L=0,1\; (E_{0+},M_{1-},E_{1+},M_{1+})$   & $\dagger$  &    &  TPWA at 1 angle not possible\\
  & full set of 4 transverse                 &   & 8(5)2 & $R_T^{00}[2],R_T^{0y}[2],R_T^{y'0}[2],R_{TT}^{00}[1],R_{TT'}^{0x}[1]$\\
  & $S,P$ wave multipoles                    &   & 8(4)3 & $R_T^{00}[3],R_{TT}^{00}[1],R_{TT}^{0x}[2],R_{TT'}^{0x}[2]$\\
 \hline
5 &$L=0,1,2\; (E_{0+},M_{1-},E_{1+},E_{2-},$ & 12(12)& 12(12)1 & $R_{T}^{00}[1],R_{T}^{0y}[1],R_{T}^{y'0}[1],R_{L}^{00}[1],R_{L}^{0y}[1],R_{L}^{x'x}[1],$\\
  &   $\;L_{0+},L_{1-})$                     &       &         & $\; R_{LT}^{00}[1],R_{LT}^{0z}[1],R_{LT}^{x'0}[1],R_{TT}^{00}[1],R_{TT'}^{00}[1],R_{TT'}^{0x}[1]$\\
  & set of 6 $S,P,D$ wave multipoles         &       & 12(5)3  & $R_{T}^{00}[3],R_{T}^{0y}[2],R_{LT}^{00}[2],R_{LT}^{0y}[3],R_{LT'}^{00}[2]$\\
 \hline
6 &$L=0,1\; (E_{0+},M_{1-},E_{1+},M_{1+},$   & $\dagger$  &    &  TPWA at 1 angle not possible\\
  &   $\;L_{0+},L_{1-},L_{1+})$              &   &14(7)2 & $R_{T}^{00}[2],R_{T}^{0y}[2],R_{T}^{x'x}[2],R_{L}^{00}[2],R_{L}^{0y}[2],R_{LT}^{00}[2],$\\
  & full set of 7 $S,P$ wave multipoles      &   &       & $\; R_{LT}^{0x}[2]$\\
  &                                          &   &14(6)3 & $R_{T}^{00}[3],R_{T}^{0y}[2],R_{LT}^{00}[2],R_{LT}^{0x}[3],R_{LT'}^{00}[2],R_{LT'}^{0x}[2]$\\
 \hline
\end{tabular}
\end{table*}

\subsection{\boldmath Comparing CEA and TPWA beyond $J=1/2$}

In Set 3 of Table~\ref{tab3}, we study a purely longitudinal model,
with two complex helicity $(H_5,H_6)$ or transversity amplitudes
$(b_5,b_6)$, four possible polarization observables, see
Table~\ref{tab:obs1} and $2L+1$ complex multipoles
$L_{\ell\pm}$. With all four observables, a CEA is possible and can
determine the two complex amplitudes up to a phase. But a TPWA with
three multipoles requires six measurements and is therefore not
possible at a single angle. However, we find a solution with four
observables at maximally two angles, and also with a minimal number
of three observables, measured at maximally three angles, a solution
exists.

Set 4 is identical to the photoproduction case. Here only electric
and magnetic multipoles contribute, and as discussed in our previous
paper~\cite{TPWA} a TPWA at a single angle is not possible. This set
can be uniquely resolved with only four observables requiring only
beam and target polarization:
$R_T^{00}[3],R_{TT}^{00}[1],R_{TT}^{0x}[2],R_{TT'}^{0x}[2]$, which
are identical to the photoproduction observables
$I[3]\,,\check{\Sigma}[1]\,,\check{H}[2]\,,\check{F}[2]$.

In Set 5, we discuss a model with six multipoles and six
non-vanishing amplitudes. In this case the CEA and TPWA are
equivalent and both can be resolved with the same number of 12
observables measured at a single angle. Again, when the information
from more than one angle is available, the number of observables can
be drastically reduced to only five, which need to be measured at
maximally three angles.

Finally, in Set 6, we discuss the full set of seven $S,P$ wave
multipoles, which requires 14 measurements for a unique solution. In
this case we find a minimal number of six observables, where again
recoil polarization can be completely avoided. A similar set is also
possible that completely avoids target polarization. With a total
number of 36 observables, a huge number of possibilities exist that
could be used to resolve all ambiguities.

The results of Set 6 with 14 measurements of six observables and two
angles for $L=1$ can be generalized theoretically for
arbitrary $L$, as was found in
photoproduction~\cite{omel,wunder,TPWA}. For each additional angular
momentum, $\ell$, each observable obtains two more Legendre
coefficients, and therefore allows for two additional independent
angular measurements. The number of multipoles increases with
$6L+1$ and the number of different measurements by
$n=12L+2$. With six observables, the number of
measurements increases by 12 for each additional angular momentum,
therefore there is no principal limit for $L$. In
practice this is, however, very different. Our present numerical
simulations are approaching a limit for $L=3$. All
examples with $L=1$ are calculated with the Mathematica
NSolve function, giving exact solutions within integer algebra. This
approach was no longer successful for $L=2$, therefore,
instead of finding exact solutions, we have done a minimization of
the coupled equations using the Mathematica NMinimize function and
random search methods. This worked very well and for the solutions
with $L=2$ the squared numerical deviation was found to
be of the order $10^{-20}$, in agreement with our work on
photoproduction.

\subsection{TPWA without Rosenbluth separation}

So far, we have always assumed that a complete separation of all
observables (response functions) of Eq.~(\ref{eq:dsigmafull}) has
been obtained in a first preparatory step. For most of these, e.g.
with $\phi$ dependence or beam polarization $h$, this is
straightforward and has been applied very successfully in the past.
A problem is the so-called Rosenbluth separation between $R_T$ and
$R_L$, which is experimentally very challenging and has only been
done in a very few cases~\cite{Blomqvist:1997qv,Defurne:2016eiy}.
However, for a TPWA the combination
$R^{\beta,\alpha}_T+\varepsilon_L R^{\beta,\alpha}_L$ can be used
and a separation is not necessary. In many cases that are discussed
in Table~\ref{tab3}, the observables $R^{\beta,\alpha}_T$ can be
replaced by the Rosenbluth combinations
\begin{equation}
R_{RB}^{\beta,\alpha} = R^{\beta,\alpha}_T+\varepsilon_L
R^{\beta,\alpha}_L\,,
\end{equation}
and we find a unique solution for all included partial waves. In the
special case of Set 1, with only three observables, this is not
possible and a fourth observable is needed.

In 2005, the Hall A Collaboration at JLab published a measurement on
`Recoil Polarization for $\Delta$ Excitation in Pion
Electroproduction', where 14 separated response functions plus two
Rosenbluth combinations had been observed in full angular
distributions at $W=1.23$~GeV and
$Q^2=1.0$~(GeV/c)$^2$~\cite{Kelly:2005jj}. In our notation, these
are
\begin{eqnarray}
&&R_{RB}^{00},\;R_{RB}^{y'0},\\
&&R_{TT}^{00},\;R_{TT}^{x'0},\;R_{TT}^{y'0},\;R_{TT}^{z'0},\nonumber\\
&&R_{LT}^{00},\;R_{LT}^{x'0},\;R_{LT}^{y'0},\;R_{LT}^{z'0},\nonumber\\
&&R_{LT'}^{00},\;R_{LT'}^{x'0},\;R_{LT'}^{y'0},\;R_{LT'}^{z'0}\,,\nonumber\\
&&R_{TT'}^{x'0},\;R_{TT'}^{z'0}\,.\nonumber
\end{eqnarray}
For a CEA, this set of observables is not complete. A complete
experiment analysis for electroproduction needs a minimum of 12
observables including both target and recoil polarization. In
fact, with two more observables involving also target polarization, a
CEA would be possible. These are e.g. $R_{LT}^{0x},R_{LT}^{0z}$ or
$R_{TT}^{0x},R_{TT}^{0z}$ or $R_{LT}^{0x},R_{TT}^{0x}$ or many
other combinations.

For a TPWA, however, the 16 observables from the Hall A experiment
are by far complete. Only a subset of 6 observables, at maximally 3
angles, is needed for a unique solution of all $S,P$ wave multipoles,
e.g.
$R_{RB}^{00}[3],R_{RB}^{y'0}[2],R_{LT}^{00}[2],R_{LT}^{x'0}[2],R_{LT'}^{00}[2],R_{LT'}^{x'0}[3]$.

\section{Conclusions}

We have explored the CEA and TPWA approaches to pseudoscalar-meson
electroproduction, extending our previous study of photoproduction.
Simple examples, corresponding to a low angular momentum cutoff,
simplify the discussion and allow one to see how the CEA and TPWA
are related. As in photoproduction, the TPWA can be accomplished
with fewer observable types supplemented by additional angular
measurements. The resulting TPWA (multipole) amplitudes have an
undetermined phase depending on energy while the CEA (transversity
or helicity) amplitudes are found with an unknown overall phase
depending on both energy and angle. Comparisons are given for
representative cases in Table~\ref{tab3}.

The CEA requires measurements involving both polarized targets and
recoil polarization, as was stressed in the study of
Ref.~\cite{DDG}. This is similar to the finding, for CEA analyses
and photoproduction, that measurements are required from two out of
the three groups containing beam-target, beam-recoil, and
target-recoil observables. Triple polarization experiments give no
further information in photoproduction, which is different from
electroproduction.  For purely transverse observables it is the
same, but for purely longitudinal $L$ and longitudinal-transverse
interference terms $LT$ and $LT'$ this is different. Already the
terms without target and recoil polarization, $R_{L}^{00}$,
$R_{LT}^{00}$ and $R_{LT'}^{00}$ have to be counted as single beam
polarizations with a polarized virtual photon. By this way of
counting, there are six triple polarization observables, see
Table~\ref{tab:obs1}, all of which can be measured in an alternative
triple polarization measurement. In electroproduction, as in
photoproduction, all 36 observables can be measured in an
alternative way, giving in total 72 possibilities for allowed
measurements. However, as was found in Ref.~\cite{TPWA}, the TPWA
can be accomplished without involving observables having both
polarized targets and recoil polarization. This is not the case for
a CEA, where at least 2 observables have to be chosen from another
group. This finding from photoproduction carries over to
electroproduction without further modification.

The present formalism can be immediately applied to data. In fact,
there exists a dataset~\cite{Kelly:2005jj} which measured 16
observables, mostly with recoil polarization but was conducted
without a polarized target. Even though this set was not complete
for a CEA, it was by far enough to fulfill the requirements of a
complete TPWA.

\begin{acknowledgments}
The work of HH and RW was supported in part by the U.S. Department
of Energy Grant DE-SC0016582. The work of LT and YW was supported by
the Deutsche Forschungsgemeinschaft (SFB 1044 and SFB/TR16).
\end{acknowledgments}

%\vfill \eject

%\eject

\end{document}